\documentclass[aps,prb,twocolumn,showpacs,amsmath,floatfix,superscriptaddress,subeqn]{revtex4-1}

\usepackage[]{graphicx}
\usepackage{subfigure}
\usepackage{bm}
\usepackage{longtable}
\usepackage{booktabs}
\usepackage{array}
\usepackage{dcolumn}
\usepackage{bm}
\usepackage[dvipdfm,colorlinks=true,citecolor=blue,linkcolor=blue,urlcolor=blue,bookmarks=true]{hyperref}
\usepackage[sort&compress]{natbib}
\newcommand{\cso}{CuSe$_2$O$_5$ }
\newcommand{\csos}{CuSe$_2$O$_5$}
\newcommand{\fref}[1]{Fig.~\ref{#1}}
\newcommand{\eref}[1]{(\ref{#1})}
\renewcommand{\eqref}[1]{Eq.~(\ref{#1})}

\usepackage{color}
\definecolor{red}{rgb}{1,0,0}

\begin{document}

\preprint{APS/123-QED}

\title{\texorpdfstring{Magnetic order and low-energy excitations in the quasi-one-dimensional antiferromagnet CuSe$_2$O$_5$ with staggered fields}{Magnetic order and low-energy excitations in the quasi-one-dimensional antiferromagnet CuSe2O5 with staggered fields}}
\author{M. Herak}\email{mirta.herak@ijs.si, mirta@ifs.hr}
\affiliation{Jo\v{z}ef Stefan Institute, Jamova 39, SI-1000 Ljubljana, Slovenia} \affiliation{Institute of physics, Bijeni\v{c}ka c. 46, HR-10000, Zagreb, Croatia}
\author{A. Zorko}
\affiliation{Jo\v{z}ef Stefan Institute, Jamova 39, SI-1000 Ljubljana, Slovenia}
\affiliation{EN-FIST Centre of Excellence, Dunajska 156, SI-1000 Ljubljana, Slovenia}

\author{M. Pregelj}
\affiliation{Jo\v{z}ef Stefan Institute, Jamova 39, SI-1000 Ljubljana, Slovenia}

\author{O. Zaharko}
\affiliation{Laboratory for Neutron Scattering, Paul Scherrer Institute, CH-5232 Villigen, Switzerland}

\author{G. Posnjak}
\affiliation{Jo\v{z}ef Stefan Institute, Jamova 39, SI-1000 Ljubljana, Slovenia}

\author{Z. Jagli\v{c}i\'{c}}
\affiliation{Institute of Mathematics, Physics and Mechanics, 1000 Ljubljana, Slovenia}

\author{A. Poto\v{c}nik}
\affiliation{Jo\v{z}ef Stefan Institute, Jamova 39, SI-1000 Ljubljana, Slovenia}

\author{H. Luetkens}
\affiliation{Laboratory for Muon Spin Spectroscopy, Paul Scherrer Institute, CH-5232 Villigen, Switzerland}

\author{J. van Tol}
\affiliation{National High Magnetic Field Laboratory, Florida State University, Tallahassee, Florida 32310, USA}

\author{A. Ozarowski}
\affiliation{National High Magnetic Field Laboratory, Florida State University, Tallahassee, Florida 32310, USA}

\author{H. Berger}
\affiliation{Institute of Physics of Complex Matter, EPFL, 1015 Lausanne, Switzerland}

\author{D. Ar\v{c}on}
\affiliation{Jo\v{z}ef Stefan Institute, Jamova 39, SI-1000 Ljubljana, Slovenia}
\affiliation{Faculty of Mathematics and Physics, University of Ljubljana, Jadranska 19, 1000 Ljubljana, Slovenia}
\date{\today}
%

\begin{abstract}
Ground state and low-energy excitations of the quasi-one-dimensional antiferromagnet \cso were experimentally studied using bulk magnetization, neutron diffraction, muon spin relaxation and antiferromagnetic resonance measurements. Finite interchain interactions promote long-range antiferromagnetic order below $T_N=17$~K. The derived spin canted  structure is characterized by the magnetic propagation vector $\mathbf{k}=(1,\:0,\:0)$ and the reduced magnetic moment $\bm{m}=[0.13(7),\:0.50(1),\;0.00(8)]\mu_B$. The values of the magnetic anisotropies determined from the field and angular dependencies of the antiferromagnetic resonance comply well with a previous electron paramagnetic resonance study and correctly account for the observed magnetic ground state and spin-flop transition.
\end{abstract}

%
%
\maketitle

\section{\label{sec:intro}Introduction}
\begin{figure}[tb]
	\centering
		\includegraphics[clip,width=\linewidth]{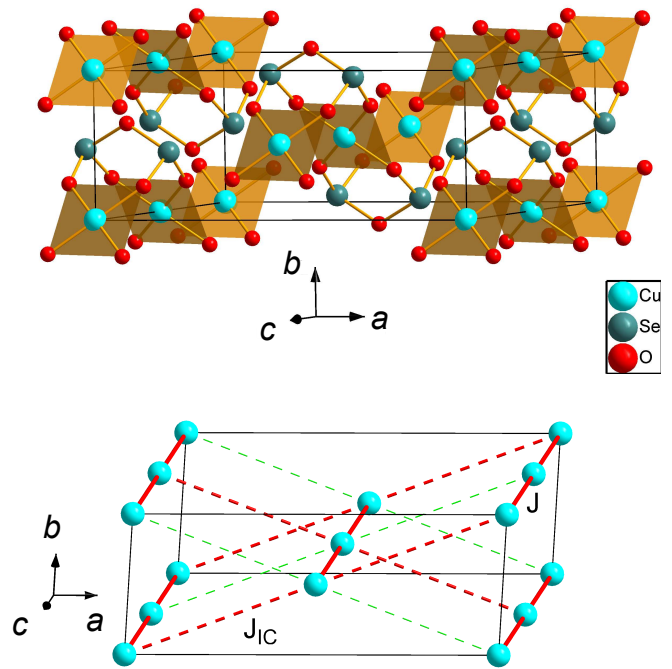}
		\caption{(Color online) Crystal structure of \csos. Lower panel shows dominant interaction paths. Solid line represents the intrachain interaction $J$, red (dark) dashed line the dominant interchain interaction $J_{IC}$ and green (light) dashed line the weak  interchain interaction $J_{IC,weak}\lesssim 0.1 J_{IC}$ according to Ref.  \onlinecite{Janson-09}.}
	\label{fig1}
\end{figure}

\indent Experimentally obtained in-depth information about the magnetic properties of one-dimensional (1D) quantum spin systems are important for testing predictions of ground states and low-energy excitations in advanced quantum-mechanical theories. A much studied representative of these systems is the spin $S=1/2$ 1D Heisenberg antiferromagnet (1D HAF) described by the simple isotropic Heisenberg Hamiltonian
\begin{equation}\label{eq:HeisenbergH}
	\mathcal{H}=J \sum\limits_{i} \bm{S}_i \cdot \bm{S}_{i+1}\;,
\end{equation}
where $i$ is the site index in the chain, and $J$ is the intrachain exchange coupling between the spins. The ground state of the $S=1/2$ 1D HAF is a Tomonaga-Luttinger liquid (TLL) and the excitations are free spinons which carry spin $1/2$ and can be created only in pairs.\cite{Faddeev-81} The excitation spectrum of the $S=1/2$ 1D HAF differs from a dispersion predicted by the classical spin-wave theory and is characterized by a continuum in the energy-momentum space,\cite{Muller-81} which was observed experimentally in several 1D spin systems.\cite{Arai-96,Dender-96,Kenzelmann-04,Lake-05} In real materials a finite interchain interaction, $J_{IC}$, always exists, and can lead to long-range order (LRO) below a finite temperature $T_N$. When $J_{IC}$ is small compared to $J$, a system is called quasi-one-dimensional (quasi-1D). In the ordered state at $T<T_N$ the excitation spectrum of these systems at low energies is usually described by a spin-wave theory,  as for an ordinary 3D antiferromagnet. At higher energies, however, where the chains start to decouple, a continuous excitation spectrum typical for a 1D system was observed.\cite{Lake-00,Lake-05,Lake-nmat05,Zheludev-00,Zheludev-01,Zheludev-02} A crossover regime from 3D LRO to 1D TLL thus exists in these systems.\cite{Lake-nmat05}\\
\indent When the crystal symmetry of quasi-1D spin systems is sufficiently low, a staggered $g$ tensor and/or a Dzyaloshinskii-Moriya interaction (DMI)\cite{MoriyaPRL-60,*Moriya-60} can be present. The combined action of staggered $g$ tensor and DMI leads to a staggered field in a finite applied field, opening a gap in the excitation spectrum of the quasi-1D chain. In the AFM ordered state, both may lead to a non-collinear staggering of the ordered moments.\cite{Dender-97,OA-PRL97,*OA-PRB99,*OA-PRB00,Kenzelmann-04}
The problem of weakly coupled $S=1/2$ HAF chains in a staggered magnetic field is still not sufficiently understood.\cite{Sato-04,Xi-11} 
For instance, if a staggered $g$ tensor and DMI are present, the staggered field can compete with the arrangement of spins favored by the interchain interaction.\cite{Sato-04,Xi-11}
The magnetic arrangement, resulting from such a competition, was recently studied by nuclear magnetic resonance (NMR) in BaCu$_2$Si$_2$O$_7$ where the presence of staggered fields was argued to cause unusual spin reorientations.\cite{Casola-12}\\
\indent \cso is a novel Cu$^{2+}$ ($S=1/2$) quasi-1D HAF, which crystallizes in a monoclinic unit cell that belongs to the $C2/c$ space group (\fref{fig1}).\cite{Meunier-76,*Becker-06} The alternating CuO$_4$ plaquettes form chains running along the $c$ crystallographic axis.
The 1D nature of this system was argued from the temperature dependence of the magnetic susceptibility which was satisfactorily described by the Bonner-Fisher curve\cite{BF-64,*Johnston-00} for $S=1/2$ and $J/k_B=157$~K.\cite{Janson-09} Band structure calculations lead to similar $J/k_B= 165$~K and reveal that the intrachain interaction is realized through a double Cu-O-Se-O-Cu path (upper panel in \fref{fig1}) with only one dominant interchain coupling of $J_{IC}/k_B=20$~K, shown in the lower panel of \fref{fig1} by dashed red (dark) lines.\cite{Janson-09} Raman scattering measurements indicated that the spin-spin correlations set in below $T\approx 100$~K,\cite{Choi-11}
which coincides with the maximum in the magnetic susceptibility.
Contrary to what is expected for a quasi-1D HAF, the temperature dependence of the magnetic specific heat extracted from the Raman scattering intensity exhibits no maximum, implying the presence of classical spin dynamics originating from the moderate interchain interactions.\cite{Choi-11} On the other hand, the recent analysis of the electron spin resonance (ESR) linewidth in the paramagnetic state suggests that at temperatures $T\gtrsim J_{IC}/k_B$ \cso essentially behaves as a 1D antiferromagnet.\cite{Herak-11} \\
\indent The low crystal symmetry and the alternating arrangement of CuO$_4$ plaquettes (\fref{fig1}) allow for staggered $g$-tensor and DMI with a DM vector confined in the $a^*c$ plane.\cite{MoriyaPRL-60,*Moriya-60} The early ESR study \cite{Herak-11} indeed confirmed the presence of staggered fields and suggested an additional symmetric anisotropic interaction. Bulk magnetic-susceptibility-anisotropy measurements showed significant deviation from the 1D HAF model, which were successfully explained\cite{Herak-11} by the extension of $S=1/2$ 1D HAF to include staggered fields.\cite{OA-PRL97,*OA-PRB99,*OA-PRB00} At $T_N=17$~K \cso undergoes a phase transition to a LRO magnetic state\cite{Janson-09, Herak-11} with so far yet unknown magnetic structure. An important question that arises from these observations is how the coexistence of the staggered field and the symmetric anisotropic exchange affects the magnetic LRO in the presence of $J_{IC}$.  We thus decided to perform a detailed experimental study of the magnetically ordered state of \cso by employing bulk magnetic  measurements, neutron diffraction, muon spin relaxation and antiferromagnetic resonance measurements.
\section{\label{sec:exp}Experimental details}
\indent The single crystalline \cso samples were synthesized by a standard chemical vapor transport method, as described previously, and characterized by X-ray diffraction.\cite{Meunier-76,*Becker-06} The samples had a platelet shape, elongated along the crystallographic $c$ axis and with the $a^*$ axis perpendicular to the platelet.\\
\indent The dc magnetic measurements were performed with a Quantum Design SQUID magnetometer in static magnetic fields ranging from 1~kOe to 50~kOe in the temperature interval between 2~K and 300~K. The measurements were performed with the magnetic field applied along the crystallographic $a^*$, $b$ and $c$ axes. The mass of the sample was $(2.9\pm0.2)$~mg. Magnetic torque measurements were performed on a home-built torque magnetometer at $4.2$~K in magnetic fields up to 8~kOe.\\
\indent Neutron diffraction on a single crystal was performed on the TriCS instrument at SINQ, Switzerland, with neutron wavelength $\lambda=2.316$~\AA. The mass of the crystal was around 10~mg. The single crystal was mounted in a CCR cooling device at a four-circle cradle. The data sets were collected at 6~K and 20~K. Neutron diffraction in magnetic field was measured on the same crystal oriented with the $b$ axis vertical in the Oxford vertical cryomagnet.\\
\indent The muon spin relaxation ($\mu$SR) experiments were conducted on the General Purpose Surface-Muon (GPS) instrument at the Swiss Muon Source (S$\mu$S), Paul Scherrer Institute (PSI), Switzerland, in the temperature range between 1.8 and 25~K in zero applied magnetic field. Measurements on powders (500~mg) were performed in a longitudinal muon polarization mode -- muon spins were polarized almost parallel ($\alpha \sim 10^{\circ}$) to the beam ($z$) direction. Measurements on a single crystal ($10\times4\times0.5$ mm$^3$) were conducted  in a transverse muon polarization mode -- the muon polarization was rotated from the $z$ direction by $\alpha \sim \pi/4$ towards $y$ direction. The asymmetry of detected positrons, emitted after muon decays, was measured with two sets of detectors; in the backward-forward ($z$) direction and in the up-down ($y$) direction. The initial asymmetry and the tilt angle $\alpha$ of the initial muon polarization were calibrated for each experiment at 25~K; i.e., well above the ordering temperature $T_N=17$~K, in the weak transverse magnetic field of 30~Oe applied in $x$ direction. Measurements were performed in veto mode, leading to negligible background signal in the case of the powder sample due to its large mass. The background signal could, however, not be avoided in the case of the single crystal due to the small thickness of the sample.\\
\indent The antiferromagnetic resonance (AFMR) measurements were performed on single-crystalline samples at $T=5$~K. Measurements in X- (9.7~GHz) and in Q- (35~GHz) band were performed on a commercial Bruker spectrometers at the Jo\v{z}ef Stefan Institute in Ljubljana. AFMR measurements at frequencies from 50~GHz to 450~GHz were performed using a custom-made transmission type spectrometers at the National High Magnetic Field Laboratory (NHMFL) in Tallahassee, Florida.\cite{HFESR} 
\section{\label{sec:results}Results}
\subsection{\label{sec:magnetic}Magnetization measurements}
\begin{figure}[tb]
	\centering
		\includegraphics[clip,width=\linewidth]{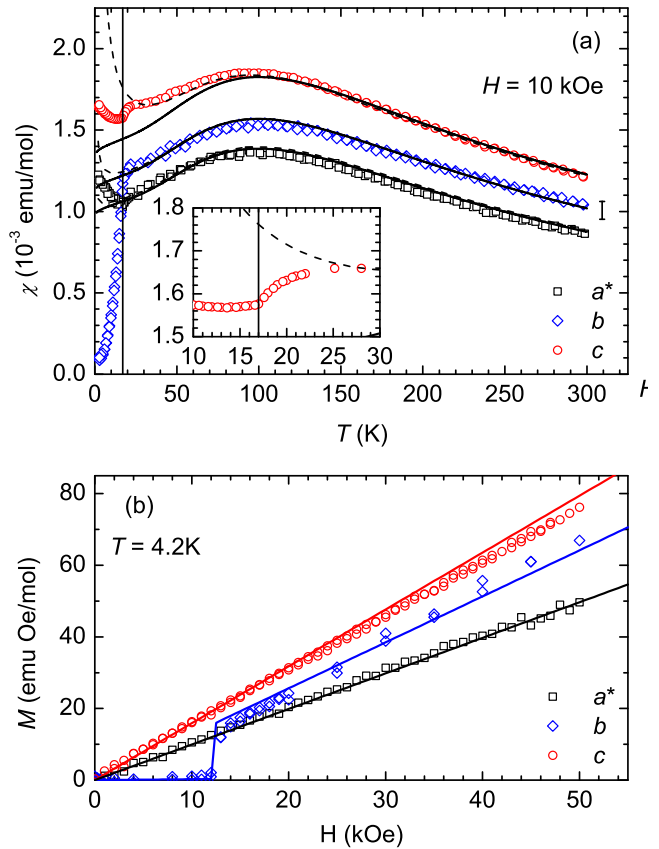}
	\caption{(Color online) (a) Temperature dependence of magnetic susceptibility measured in the field $H=10$~kOe applied along $a^*$, $b$ and $c$ axis. Solid lines represent fits to the $S=1/2$ 1D HAF model with $J=156$~K, while dashed line also includes staggered susceptibility (see text). Error bar resulting from the uncertainty in the mass of the sample is shown on the side. Vertical line represent $T_N=17$~K. Inset: Expanded region around $T_N$ showing the disagreement between the experimental data (red circles) and the model (dashed line) for $c|| H$. (b) Field dependence of magnetization at $T=4.2$~K. Solid lines represent the results of calculations rescaled to match the observed values (see Sec. \ref{sec:AFMR}).}
	\label{fig2}
\end{figure}

\indent Temperature dependence of dc magnetic susceptibility ($\chi=M/H$, where $M$ is the sample magnetization) measured in $H=10$~kOe applied along $a^*$, $b$ and $c$ axes is shown in \fref{fig2}(a). In the paramagnetic state the measured data can be described by the $S=1/2$ 1D HAF model\cite{Johnston-00} with $J/k_B=156$~K, if the $g$-factor values measured by ESR\cite{Herak-11} ($g_{a^*}=2.064$, $g_b=2.140$, $g_c=2.226$) are taken into account [solid lines in \fref{fig2}(a)]. We stress, however, that for the $c$ direction a disagreement with the model starts already below $T\approx T_{max}$, i.e., far above the transition to the magnetic LRO state. The presence of the staggered DMI and the staggered $g$ tensor leads to finite staggered field $h$ proportional to the applied field, $h_i=c_{s,i}\:H$ ($i=a^*,\,b,\,c$),\cite{OA-PRL97,OA-PRB99,*OA-PRB00} which results in the anisotropic staggered susceptibility for $T<J/k_B$, $\chi_{s,i}\propto c_{s,i}^2$.\cite{OA-PRL97,OA-PRB99,*OA-PRB00} The temperature dependence of the staggered susceptibility reflects in the Curie-like term $\chi \propto 1/T$, which strongly varies with the direction of the magnetic field, as has been observed for the 1D $S=1/2$ system [PM·Cu(NO$_3$)$_2$·(H$_2$O)$_2$]$_n$ (PM = pyrimidine).\cite{Feyerherm-00} Indeed the measured magnetic susceptibility $\chi_c$ can be satisfactorily described if the staggered susceptibility with coefficient $c_{s,c}=0.17$ is added to the 1D HAF model [dashed line in \fref{fig2}(a)], in rather good agreement with the previous anisotropy results, $c_s=0.13$.\cite{Herak-11} For the $a^*$ and $b$ direction the data are well described by the 1D HAF model combined with staggered susceptibility using the previously obtained $c_{a^*}$ and $c_b$ [see \fref{fig2}(a)]. Below $\approx 22$~K there is a drastic disagreement between the data and the model even if the staggered susceptibility is included [inset in \fref{fig2}(a)]. This was also observed in previous anisotropy measurement, however, no satisfactory explanation for this behavior can be given at the moment.\\
\indent On cooling below $T_N$, $\chi$ measured along the $b$ axis decreases and saturates at the value of $9\cdot 10^{-5}$~emu/mol below 4~K, while $\chi_{a^*}$ and $\chi_c$ slightly increase. This suggests an almost collinear spin arrangement with the $b$ axis as the easy axis, which is in agreement with the magnetization measurements at $4.2$~K, shown in \fref{fig2}(b), where a spin-flop (SF) transition is observed for the field $H_{SF}\approx13$~kOe applied along the $b$ axis. In contrast, the magnetization changes linearly with field for $a^*$ and $c$ directions up to the highest applied field of 50~kOe. 
\subsection{\label{sec:neutrons}Neutron diffraction measurements}
\begin{figure}[tb]
	\centering
		\includegraphics[clip,width=1.00\linewidth]{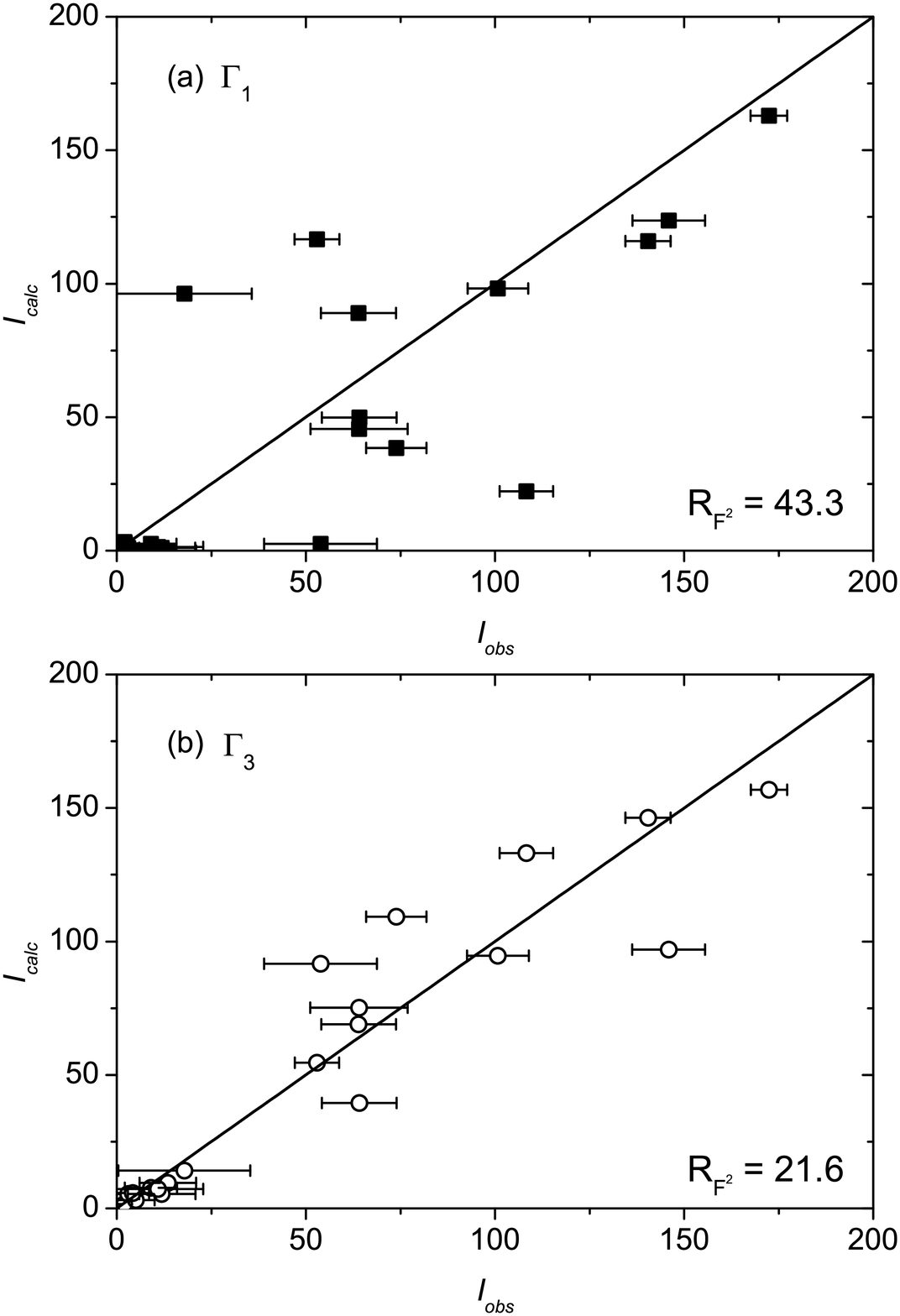}
	\caption{Agreement of calculated and observed intensities of magnetic reflections for models (a) $\Gamma_1$ and (b) $\Gamma_3$ (see text).}
	\label{fig3}
\end{figure}

\indent Neutron diffraction is a powerful tool for determining the magnetic order. Therefore, we have employed it to determine the magnetic structure of \cso more precisely. A refinement of the crystal structure at 6~K confirms the room-temperature structural model published previously\cite{Meunier-76,*Becker-06} but with slightly different values of the cell parameters; $a=12.30(2)$\AA, $b=4.89(2)$\AA, $c=7.88(1)$\AA~ and $\beta=112.24(15) ^{\circ}$. Below $T_N=17$~K new reflections of magnetic origin appear. These correspond to the magnetic propagation vector $\mathbf{k}=(1,\:0,\:0)$ and are summarized in Table \ref{tab:neutrint} given in Appendix \ref{appA}.\\
\indent In order to determine the magnetic structure, we start with a representation analysis using the {\sc basireps} program.\cite{Rodriquez-Carvajal} The magnetic moment in \cso originates from the Cu$^{2+}$ ions at the $4a$ Wyckoff site. The two possible irreducible representations, which connect the Cu-sites related by a twofold screw axis, are given in Table \ref{tab:irreps}. The best agreement with the experimental data is obtained for the irreducible representation $\Gamma_3$ with $\chi^2=5.03$ and $R_{F^2}=21.6$. The representation $\Gamma_1$ can be discarded due to much poorer agreement with the experimental data,  $\chi^2=21.1$ and $R_{F^2}=43.3$. The agreement between the observed and calculated intensities for those two models is shown in \fref{fig3}.\\ 
\begin{table}[tb]
	\centering
	\caption{Irreducible representations $\Gamma_1$ and $\Gamma_3$ of the little group for $\mathbf{k}=(1,\:0,\:0)$ in the space group $C2/c$.}
		\begin{tabular*}{\linewidth}{@{\extracolsep{\fill}} c | c c }
   \hline \hline
   	 & $\Gamma_1$ & $\Gamma_3$\\
   	 \hline
   	 $x$, $y$, $z$ & $(u, v, w)$ & $u, v, w$\\
   	 $-x$, $y$, $-z+1/2$ & $(-u, v, w)$		 & $(u, -v, w)$\\
   	 \hline \hline
		\end{tabular*}
			\label{tab:irreps}
\end{table}

\begin{figure}[tb]
	\centering
		\includegraphics[clip,width=0.9\linewidth]{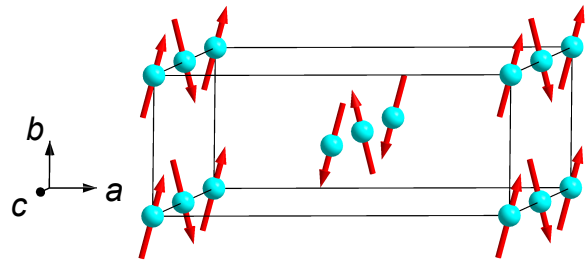}
		\caption{(Color online) Proposed zero-field magnetic structure in the ordered state of \cso obtained from neutron diffraction.}
	\label{fig4}
\end{figure}

The components of the magnetic moment were obtained from the data refinement using the {\sc fullprof} program.\cite{fullprof} The magnetic moment of the Cu$^{2+}$ ion at the crystallographic position $(0,0,0)$ in the crystallographic $(abc)$ coordinate system is $\mathbf{m}=(m_a, m_b, m_c) =[0.13(7), \: 0.50(1), \: 0.00(8)]\mu_B$. The value of the magnetic moment $|m|=0.52(2)\mu_B$ is thus significantly smaller than the full magnetic moment of $1\mu_B$ for $S=1/2$. The refined magnetic structure is shown in \fref{fig4}. The magnetic moments on a chain at the positions $(0,0,0)$ and $\left(0,0, \frac{1}{2}\right)$ possess different $b$ components in the representation $\Gamma_3$, but equal $a$-components, which results in finite magnetization on each chain. On the other hand, the magnetic moments at the positions $\left(\frac{1}{2}, \frac{1}{2}, 0\right)$ and $\left(\frac{1}{2}, \frac{1}{2}, \frac{1}{2}\right)$ on the neighboring chain have the same symmetry as the first two, but are antiferromagnetically coupled to them so the total magnetic moment in the unit cell is zero.\\ 
\indent Finally, we have measured two magnetic reflections in the magnetic field $H=40$~kOe applied along the $b$ axis to test the magnetic structure above the SF field. The comparison of the integrated intensities for these reflections in zero field and $H=40$~kOe is given in Table \ref{tab:neutronH}. Although only two reflections were measured, information about the orientation of spins can be extracted. The best agreement between calculated and observed intensities is found for $\Gamma_1$ and $\mathbf{m}=(m_{a^*}, m_b, m_c) \approx(0.46,\; 0,\; 0.05)\mu_B$. An almost orthogonal orientation of the spins in this field with respect to zero field is expected, since the spin-flop transition was observed at $H_{SF}\approx 13$~kOe.
\begin{table}[tb]
	\centering
	\caption{Integrated intensities of magnetic reflections in $H = 0$~Oe and $H = 40$~kOe for $(1 0 \bar{1})$ and $(\bar{1} 0 \bar{1})$ reflections.}
		\begin{tabular*}{\linewidth}{@{\extracolsep{\fill}} c | c c }
   \hline \hline
   	$h\; k\; l$ & $(1 0 \bar{1})$ & $(\bar{1} 0 \bar{1})$\\
   	 \hline
   	 $H=0$~Oe & $229 \pm 10$ & $117 \pm 8$\\
   	 $H=40$~kOe & $169 \pm 12$ & $37 \pm 7$\\
   	 \hline \hline
		\end{tabular*}
			\label{tab:neutronH}
\end{table}
\subsection{\label{sec:muSR}Muon spin relaxation}
\indent For an independent proof of the refined magnetic structure, a complementary local-probe technique was used. $\mu$SR is a highly powerful method for detecting magnetism on a microscopic level.\cite{Yaouanc} The almost 100\% spin-polarized muons that stop in a sample probe the local magnetic field $B_\mu$, which leads to coherent oscillations of the muon polarization for a static field and a monotonic decay of polarization for fast fluctuations of the field. In the case of a single quasi-static magnetic field the muon polarization along the initial polarization will change in powder samples as\cite{Yaouanc}
\begin{equation}
P^z_{\rm pwd}(t,B_\mu,\lambda_L,\lambda_T) = \frac{1}{3}{\rm e}^{-\lambda_L t}+\frac{2}{3}{\rm e}^{-\lambda_T t}{\rm cos}(\gamma_\mu B_\mu t)
\label{eq:muSRPpow}
\end{equation}
where longitudinal muon relaxation $\lambda_L$ and transverse relaxation $\lambda_T$ are taken into account ($\gamma_\mu = 85.16$~kHz/G is the muon gyromagnetic ratio). The former arises from finite dynamics of the internal field whereas the letter additionally includes a distribution of local fields. The non-oscillating "$\frac{1}{3}$"-tail signal corresponds to muons being initially polarized along the internal field and is thus a fingerprint of the magnetic order, alongside the oscillating signal.\\
\begin{figure}[bt]
\includegraphics[trim = 7mm 25mm 3mm 15mm, clip, width=1\linewidth]{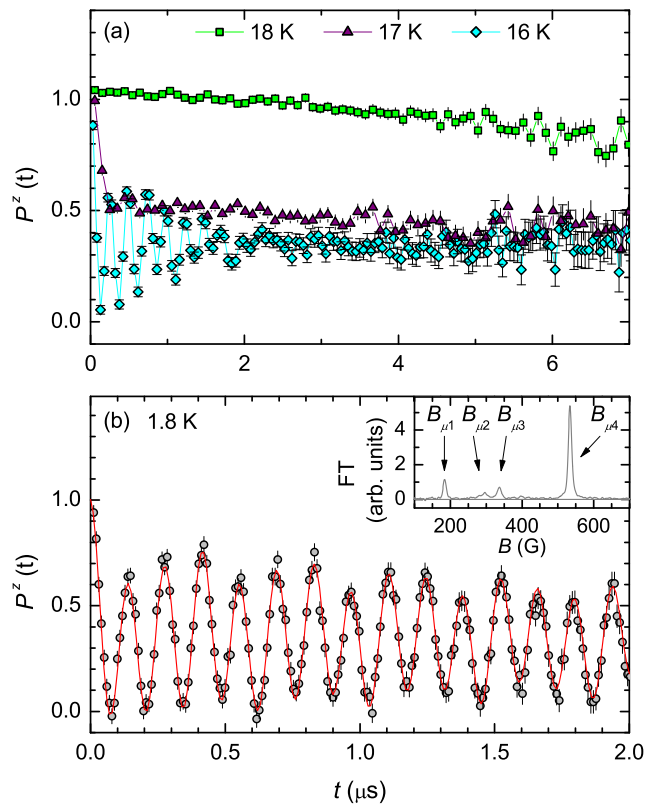}
\caption{(Color online) (a) Time dependence of muon polarization along the beam direction for longitudinal muon polarization in the \cso~powder sample at several selected temperatures close to the ordering temperature $T_N=17$~K. (b) Fit of the measured low-temperature polarization (circles) to the four-component model (solid line) given by Eq.~(\ref{eq:muSRP}). Inset: Real part of the Fourier transform of the 1.8~K dataset.}
\label{fig5}
\end{figure}
\indent The magnetic ordering in \cso~is witnessed by a clear change of the polarization curve at 17~K. Below this temperature oscillations of the polarization appear [Fig.~\ref{fig5}(a)]. The real part of a Fourier transform of the data below 17~K, which directly gives the field distribution at the muon sites, reveals four distinct components [inset in \fref{fig5}(b)]. We therefore fit the experimental polarization of the powder sample below $T_N$ with the four-component function 
\begin{equation}
P^z= \sum_{i=1}^{4}{f_i P^z_{\rm pwd}(t,B_{\mu i},\lambda_{L},\lambda_{Ti})},
\label{eq:muSRP}
\end{equation}
where $f_i$ denotes the fraction of the $i$-th component. The fitting of the 1.8~K dataset [Fig.~\ref{fig5}(b)] yields internal magnetic fields $B_{\mu 1}=185(3)$~G, $B_{\mu 2}=297(5)$~G, $B_{\mu 3}=337(3)$~G and $B_{\mu 4}=533(3)$~G, transverse relaxation rates $\lambda_{T1}=0.20(2)$~$\mu$s$^{-1}$, $\lambda_{T2}=1.0(5)$~$\mu$s$^{-1}$, $\lambda_{T3}=0.26(2)$~$\mu$s$^{-1}$ and $\lambda_{T4}=0.14(2)$~$\mu$s$^{-1}$, the longitudinal relaxation rate $\lambda_{L}=0.01$~$\mu$s$^{-1}$ and fractions $f_1=0.14(2)$, $f_2=0.14(2)$, $f_3=0.11(2)$ and $f_4=0.61(2)$. The polarization at longer times is almost time independent and approaches the $\frac{1}{3}$ value. This demonstrates that the sample is 100\% ordered and that the spin dynamics in the ground state is marginal on the muon time scale.\\ 
\indent The observation of four distinct internal magnetic fields reveals that muons stop at four different crystallographic sites, because the magnetic order does not reduce the symmetry of the crystallographic unit cell. The temperature dependence of the four internal fields and the transverse relaxation rates are shown in Fig.~\ref{fig6}. Since the fields are proportional to the ordered magnetic moment, their temperature dependence directly yield the temperature evolution of the magnetic order parameter in \csos. The relaxation rates, on the other hand, evidence an increasing spin relaxation rate when approaching the ordering temperature $T_N$, which can be either explained by an increased magnon density or by an increased width of local-field distribution\cite{Maeter-12} with increasing temperature.\\
\indent Measurements on the single crystal further allow us to set the actual direction of the internal fields at all four stopping sites. We performed these measurements at 1.8~K for two different orientations of the crystal; (a) with the $a^*$ and $b$ crystallographic axes oriented in the $z$ and $y$ directions, respectively, and (b) with the crystal rotated by $\pi/2$ along the $z$ direction so that the $c$ crystallographic axis was pointing along the $y$ direction. Due to the symmetry of the magnetic space group, for each muon stopping site $i$ four different directions of the magnetic field are allowed; $(\theta_i^1,\,\varphi_i^1)$, $(\theta_i^2,\,\varphi_i^2)$, $(\theta_i^3,\,\varphi_i^3)$, and $(\theta_i^4,\,\varphi_i^4)$. These are symmetry-related and are embedded into polarization functions $P^{y,z}_{\rm sc} (t,B_{\mu i},\theta_i,\varphi_i,\lambda_{L i},\lambda_{T i},\alpha)$ (see Appendix~\ref{appB}). For the second orientation the $\varphi_i^j$ parameters are increased by $\pi/2$ with respect to the first orientation while the $\theta_i^j$ parameters remain the same. We fitted simultaneously the powder data to Eq.~(\ref{eq:muSRP}) [see Fig.~\ref{fig5}(b)] and the three single-crystal datasets recorded for the two crystallographic orientations [see Fig.~\ref{fig7}(a,b)] to equations
\begin{widetext}
\begin{subequations}
\label{eq:muSRPsc}
\begin{eqnarray}
P^z_{a^*}&=& (1-P_{\rm bgd}) \sum_{i=1}^{4}{f_i P^z_{\rm sc}(t,B_{\mu i},\theta_i,\varphi_i,\lambda_{L},\lambda_{T i},\alpha)} + P_{\rm bgd}{\rm e}^{-\frac{(\Lambda t)^2}{2}},   \\
P^y_{b}&=& (1-P_{\rm bgd}) \sum_{i=1}^{4}{f_i P^y_{\rm sc}(t,B_{\mu i},\theta_i,\varphi_i,\lambda_{L},\lambda_{T i},\alpha)} + P_{\rm bgd}{\rm e}^{-\frac{(\Lambda t)^2}{2}},   \\
P^y_{c}&=& (1-P_{\rm bgd}) \sum_{i=1}^{4}{f_i P^y_{\rm sc}(t,B_{\mu i},\theta_i,\varphi_i+\pi/2,\lambda_{L},\lambda_{T i},\alpha)} + P_{\rm bgd}{\rm e}^{-\frac{(\Lambda t)^2}{2}},
\end{eqnarray}
\end{subequations}
\end{widetext}
where the two single-crystal polarization functions $P^{y,z}_{\rm sc}$ are given by Eq.~(\ref{eqa5}) with an additional background signal $P_{\rm bgd}{\rm e}^{-\frac{(\Lambda t)^2}{2}}$. The above-reported powder muon relaxation rates and fractions $f_i$ do not change when adding the single-crystal datasets, while the background signal in these datasets amount to 26\% ($\Lambda=0.29$~$\mu$s$^{-1}$). Such a high background is not surprising for thin single crystals. The local magnetic fields at the four muon stopping sites are summarized in Tab.~\ref{table1}.\\
\begin{figure}[tb]
\includegraphics[trim = 7mm 25mm 3mm 15mm, clip, width=1\linewidth]{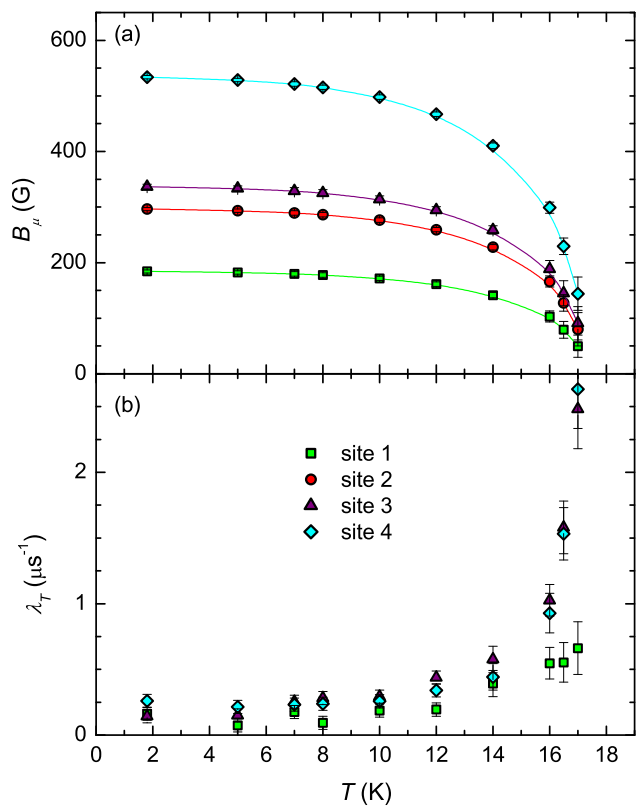}
\caption{(Color online) Temperature dependence of (a) internal fields and (b) transverse muon relaxation rate for four crystallographically nonequivalent muon stopping sites in the \cso powder sample.}
\label{fig6}
\end{figure}
\begin{table}[b]
\caption{The local magnetic field $B_{\mu i}$, the polar angle $\theta_i$ and the azimuthal angle $\varphi_i$ (the former is given with respect to the $a^*$ crystallographic axis and the latter with respect to the $c$ axis) at four muon stopping sites $i$ at 1.8~K. The corresponding component of the magnetic-field vectors ${\bf B}_{\mu i}$ in the $a^*bc$ orthogonal system are also given.}
\begin{center}
\begin{ruledtabular}
\begin{tabular}
{c c c c c} $i$ & $B_{\mu i}$ (G) & $\theta_i$ & $\varphi_i$ & ${\bf B}_{\mu i}$ (G)\\\hline 
1&185&1.22&1.97&(62,\,-159,\,67)\\
2&297&2.24&2.63&(-185,\,112,\,203)\\
3&337&2.25&2.25&(-211,\,211,\,156)\\
4&533&1.71&5.08&(-77,\,-494,\,-187)
\end{tabular}
\end{ruledtabular}
\end{center}
\label{table1}
\end{table}  
\begin{figure}[tb]
\includegraphics[trim = 7mm 25mm 3mm 15mm, clip, width=1\linewidth]{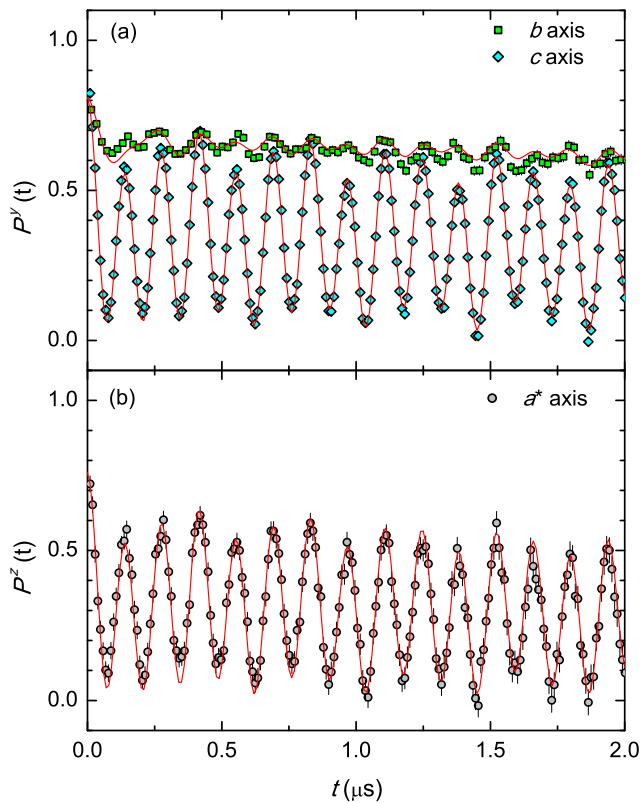}
\caption{(Color online) Simultaneous fit (solid lines) of muon polarization data (symbols) to Eqs.~(\ref{eq:muSRPsc}) along three crystallographic axes of the single-crystal \cso~sample, measured in transverse muon polarization mode (a) along and (b) perpendicular to the beam direction.}
\label{fig7}
\end{figure}
\indent Having determined the magnetic fields at the four muon stopping sites we next critically verify the magnetic order determined in the neutron diffraction experiment (\fref{fig4}). To be able to perform this assessment, the knowledge of the muon stopping sites is needed. Muons possessing positive charge are likely to stop at electrostatic-potential minima of the crystal structure, which has been shown before at several instances.\cite{Luetkens, Pregelj-12, Zorko, Maeter-09} In order to find the electrostatic-potential minima of the \cso~crystal structure, we performed density-functional-theory (DFT) calculation, using the {\sc pwscf} program of the Quantum Espresso software package.\cite{Giannozzi} A self-consistent electron density distribution was calculated, which yielded a spatial profile of the electrostatic potential. A global electrostatic-potential minimum was found at $\bm{R}_1=(0.17,0.01,0.22)$ (Wyckoff position $8f$) and three local minima at $\bm{R}_2=(0.29,0.25,0.400)$ ($8f$), $\bm{R}_3=(0,0.5,0)$ ($4b$) and $\bm{R}_4=(0,0.11,0.25)$ ($4e$).\\ 
\begin{table}[b]
\caption{The muon stopping sites $\bm{P}_i$ found at $8f$ Wyckoff positions, the corresponding dipolar magnetic fields ${\bf B}_i$, the distance $r_{\bm{P}_i-\bm{R}_j}$ to the closest electrostatic-potential minimum $\bm{R}_j$ and $r_{\bm{P}_i-{\rm O}}$ to the closest oxygen site. All the fields are within\cite{foot} 
$\sigma/B_{\mu i}=5\%$ of the experimentally determined values $B_{\mu i}$ and are given in the $a^*bc$ orthogonal system.}
\begin{center}
\begin{ruledtabular}
\begin{tabular}
{c c c c c} $i$ & $\bm{P}_i$ & ${\bf B}_i$ (G) & $r_{\bm{P}_i-\bm{R}_j}$ & $r_{\bm{P}_i-{\rm O}}$ \\\hline 
1&(0.19,\,0.01,\,0.23)&(58,\,-164,\,71)& 0.21 \AA&1.39 \AA\\
2&(0.33,\,0.40,\,0.06)&(-191,\,120,\,211)& 1.82 \AA&0.97 \AA\\
3&(0.32,\,0.44,\,0.02)&(-214,\,203,\,146)& 1.75 \AA&1.05 \AA\\
4&(0.35,\,0.49,\,0.32)&(-67,\,-512,\,-170)& 0.36 \AA&1.08 \AA
\end{tabular}
\end{ruledtabular}
\end{center}
\label{table2}
\end{table}  
\indent We further calculated the dipolar magnetic field at these sites by taking into account all spins within a sphere large enough to assure convergence of these calculations. Since \cso~is an insulator and all the potential minima are located outside the exchange paths, the dipolar contribution to the magnetic field at these sites is expected to be by far dominant. The dipolar fields at all these sites (except $\bm{R}_3$) are in the range between 189 and 654~G, thus seemingly well suiting the experimental fields $B_{\mu 1-4}=185-533$~G. However, the direction of the calculated ${\bf B}_{\rm calc}$ and experimentally determined fields ${\bf B}_{\mu i}$ are very different thus yielding large relative deviations $\frac{\sigma}{B_{\mu i}}>1$,\cite{foot}
 except for site $\bm{R}_1$, where the calculated field is the closest to the measured field ($\frac{\sigma}{B_{\mu 1}}=0.46$).\\
\indent Therefore, the muons do not seem to stop at the electrostatic-potential minima of the unperturbed \cso~structure. In order to determine the possible muon stopping sites we calculated the dipolar magnetic field on a $100\times100\times100$ mesh of the unit cell and searched for positions where the calculated fields match the experimental fields $B_{\mu 1-4}$. These positions are summarized in Tab.~\ref{table2}, where the distances to the closest electrostatic-potential minima and to the closest oxygen site are also shown.\\
\indent The position $\bm{P}_1$ corresponding to the smallest experimentally determined local field $B_{\mu 1}$ is found only 0.21~\AA~away from the global electrostatic-potential minimum at $\bm{R}_1$. A slight local modification of the electrostatic potential by the positively charged muon and/or slightly different magnetic moment ${\bf m}$ are likely reason for the small mismatch between $\bm{P}_1$ and $\bm{R}_1$. The other three sites are found further away from the minima. However, they are all positioned 1.03(5)~\AA~away from oxygen. The muon is well-known for its affinity of "bonding" to the oxygen ion with the corresponding bond length of about 1.0~\AA,\cite{Holzschuh} which is in nice agreement with our determination of the muon stopping sites. We stress that the site with the largest internal field $\bm{P}_4$ is also found 0.36~\AA~away from the global electrostatic-potential minimum. This can explain its dominant occupation $f_4=61\%$.  The accordance of the observed distances between the muon stopping sites and oxygen sites and the reported muon-oxygen "bond" length suggest that the muon perturbs the electrostatic potential of \csos. However, it preferentially remains relatively close to the global electrostatic-potential minimum of the unperturbed structure. The convincing agreement of the measured and the calculated magnetic fields at the determined muon stopping sites sets a firm confirmation on the magnetic structure determined by the neutron scattering experiment.
\subsection{\label{sec:AFMR}Antiferromagnetic resonance}
\begin{figure}[tb]
	\centering
		\includegraphics[clip,width=\linewidth]{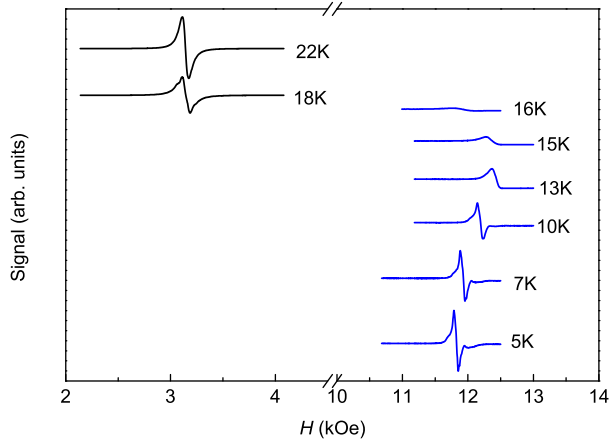}
	\caption{(Color online) The temperature dependence of the antiferromagnetic resonance line shape measured at frequency $\nu=9.7$~GHz for field applied along the $b$ axis. Paramagnetic spectra at $T>T_N$ are also shown for comparison.}
	\label{fig8}
\end{figure}

\begin{figure}[tb]
	\centering
		\includegraphics[clip,width=\linewidth]{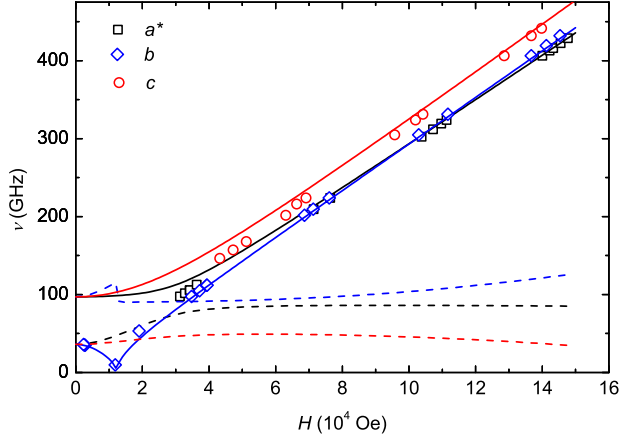}
	\caption{(Color online) The field dependence of the AFMR frequency measured at $T=5$~K. Solid and dashed lines show the results of calculations for parameters $J=157$~K, $J_{IC}=0.1\:J$, $D_a^*= -0.044$, $D_c=0.0255$, $\delta_{a^*}=0.00046$, $\delta_b=0$ and $\delta_c=-0.001$ of the Hamiltonian \eref{eq:Hamiltonian}.}
	\label{fig9}
\end{figure}

\begin{figure}[tb]
	\centering
		\includegraphics[clip,width=\linewidth]{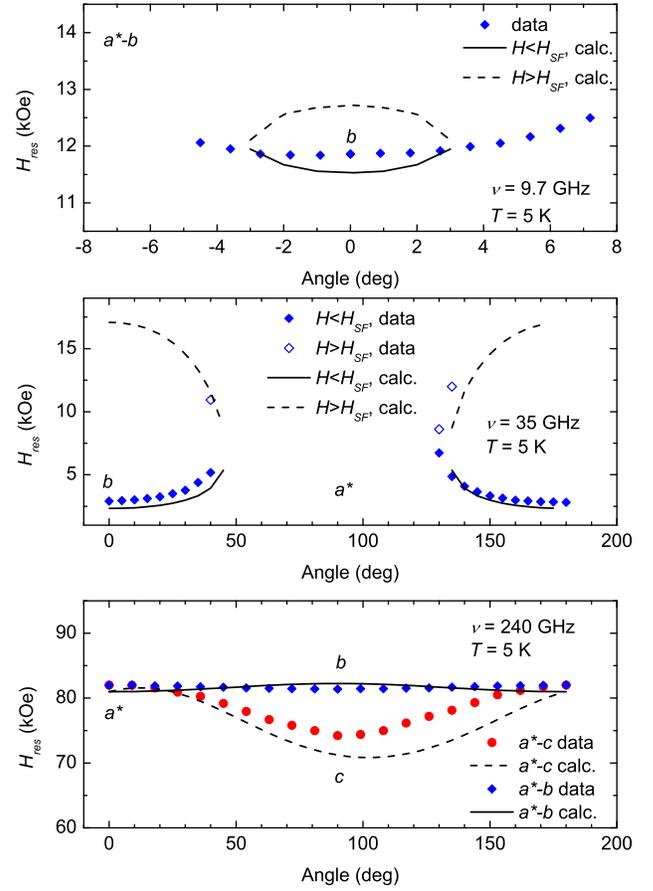}
	\caption{(Color online) The angular dependence of the AFMR resonance modes in the $a^*c$ and the $a^*b$ planes measured at different frequencies. The solid and dashed lines represent a fit to the model described in the text using parameters $J=157$~K, $J_{IC}=0.1\:J$, $D_a^*= -0.044$, $D_c=0.0255$, $\delta_{a^*}=0.00046$, $\delta_b=0$ and $\delta_c=-0.001$ in Hamiltonian \eref{eq:Hamiltonian}.}
	\label{fig10}
\end{figure}

\indent Next, we decided to perform AFMR measurements, which can provide additional information of the spin Hamiltonian responsible for the onset of the above-determined magnetic order. In the AFMR theory the magnetic order and the low-energy excitations of a spin system are described within a molecular--field approximation. The resonant frequencies of the sublattice magnetizations induced by a microwave field in the finite applied magnetic field are associated to the exchange and anisotropy molecular fields felt by the sublattice magnetizations.\cite{Kittel-51} Studying the AFMR is thus an alternative way of obtaining information about the long-range order, superexchange and magnetic anisotropy of a sample, with a high precision characteristic of magnetic-resonance experiments. Below $T_N$ the paramagnetic ESR signal quickly disappears in \cso and is replaced by a shifted temperature dependent resonances (\fref{fig8}), suggesting that these resonances belong to AFMR modes. The field dependence of the resonant frequency for $a^*$, $b$ and $c$ directions measured at $T=5$~K is shown in \fref{fig9}. Finally, we have also measured angular dependencies of the resonance field $H_{res}$ at $T=5$~K and $\nu=240$~GHz in the $a^*b$ and the $a^*c$ plane, and at $\nu=35$~GHz and $\nu=9.7$~GHz in the $a^*b$ plane. The results are summarized in \fref{fig10}. At $240$~GHz the anisotropy is much larger in the $a^*c$ plane than in the $a^*b$ plane. For X- and Q-band measurements the AFMR modes are observed only for the $b$ direction due to the experimental limitations.\\
\subsection{\label{sec:torque}Torque magnetometry}
\begin{figure}[tb]
	\centering
		\includegraphics[clip,width=1.00\linewidth]{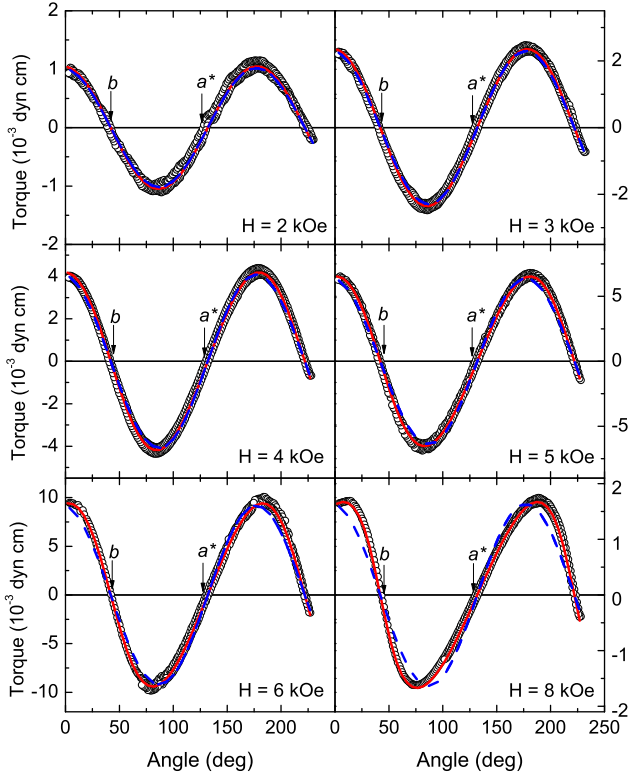}
	\caption{(Color online) Torque measured at $T=4.2$~K in $a^*b$ plane. Dashed blue line is obtained from \eqref{eq:torque} for $\chi_{a^*}$ and $\chi_b$ taken from susceptibility measurements, \fref{fig2}(a). Solid red line is the result of calculations taking into account the same parameters as in Figs. \ref{fig9} and \ref{fig10}. The values obtained by calculations are rescaled to match the observed torque amplitude in the same way as magnetization in \fref{fig2}(b).}
	\label{fig11}
\end{figure}

\indent The torque magnetometry can be a useful tool for detecting spin reorientations even in magnetic fields which are substantially smaller than the critical field of reorientation.\cite{Herak-10,*HerakSSC-11} This is because the measured magnetic torque $\bm{\Gamma} = V \bm{M} \times \bm{H}$ ($V=$ is the sample volume) is sensitive to the direction of the induced magnetization in the sample which changes when the field approaches the critical field if the direction of the field does not coincide with the easy axis (if it does then a spin flop is observed at the critical field $H_{SF}$). In the case of a uniaxial antiferromagnet in a field $H\ll H_{SF}$ the angular dependence of the measured component of torque $\Gamma_z$ for field $H$ rotating in $xy$ plane is given by
\begin{equation}\label{eq:torque}
	\Gamma_z = \dfrac{m}{2 M_{mol}}\: H^2 \Delta \chi_{xy} \sin(2\phi-2\phi_0)
\end{equation}
where $m$ is the mass, $M_{mol}$ is the molar mass, $\Delta \chi_{xy}=\chi_x - \chi_y$, $\phi$ is the goniometer angle and $\phi_0$ is the angle the $x$ axis makes with the goniometer zero angle. The torque measured at $T=4.2$~K in the $a^*b$ plane in different magnetic fields is shown in \fref{fig11}. We have also plotted the expected angular dependence (dashed blue lines), \eqref{eq:torque}, where $x=a^*$, $y=b$, $\phi_0=132^{\circ}$ and the values of $\chi_{a^*}$ and $\chi_b$ are taken from the magnetic susceptibility results at $T=4.2$~K shown in \fref{fig2}(a). Apart from the small observed hysteresis, the expected behavior, \eqref{eq:torque}, is observed in low fields. However, for $H\geq 5$~kOe the deviation from \eref{eq:torque} is observed which becomes more and more pronounced with increasing field. This type of behavior was not observed in the $a^*c$ plane, where even the torque in the highest applied field of $H=8$~kOe obeys \eqref{eq:torque}.\\
\subsection{\label{sec:Modeling}Modeling}
\begin{figure}[tb]
	\centering
		\includegraphics[width=0.15\textwidth]{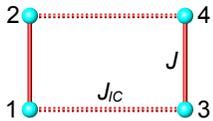}
	\caption{(Color online) Basic cell describing the interactions present in \csos. Numbers indicate the four magnetic sublattices.}
	\label{fig12}
\end{figure}

In order to describe the above presented experimental results we start with the spin Hamiltonian
\begin{eqnarray}\label{eq:Hamiltonian}
	\mathcal{H}=\sum\limits_{\textup{all}\;\textup{chains}}\mathcal{H}_{1D} + \mathcal{H}_{IC}\, ,
\end{eqnarray}
where the single-chain Hamiltonian $\mathcal{H}_{1D}$ and the interchain-interaction Hamiltonian $\mathcal{H}_{IC}$ are given by\cite{Herak-11}
%
\begin{subequations}\label{eq:Hamiltparts}
\begin{align}\nonumber
	\mathcal{H}_{1D} &= J \: \sum\limits_{i} \mathbf{S}_i \cdot \mathbf{S}_{i+1}
	+ \sum\limits_i (-1)^i \:\mathbf{D} \cdot \left( \mathbf{S}_i \times \mathbf{S}_{i+1}\right) +\\ \label{eq:H1D}
	&+  \sum\limits_{i} \mathbf{S}_i \cdot\hat{\bm{\delta}}J \cdot \mathbf{S}_{i+1} -
	\mu_B \sum\limits_{i} \mathbf{S}_i \cdot \hat{\mathbf{g}}_i \cdot \mathbf{H},\\
\label{eq:HIC}
	\mathcal{H}_{IC}& = J_{IC} \sum\limits_{<i,j>} \mathbf{S}_i \cdot \mathbf{S}_j \, .
\end{align}
\end{subequations}
The sum in \eref{eq:H1D} runs over spins on one chain, while the sum in \eref{eq:HIC} runs over spins $i$ and $j$ which reside on neighboring chains. $\mathbf{D}$ is the DM vector which, due to crystal symmetry of \csos, is restricted to\cite{MoriyaPRL-60,*Moriya-60} $\mathbf{D}=(D_{a^*}, 0, D_c)$. The DM vector is staggered which is taken into account by $(-1)^i$ in the DM term. The tensor $\hat{\bm{\delta}}$ represents the symmetric anisotropic exchange and is assumed to be diagonal in the $a^*bc$ coordinate system
\begin{equation}\label{eq:delta}
	 \hat{\bm{\delta}} = \begin{bmatrix}
       \delta_{a^*} & 0 & 0   \\[0.3em]
       0 & \delta_b & 0 \\[0.3em]
       0 & 0 & \delta_c
     \end{bmatrix}\,.
\end{equation}
The $g$ tensor in \cso is staggered, $\hat{\mathbf{g}}_i=\hat{\mathbf{g}}_u + (-1)^i \hat{\mathbf{g}}_s$, where $\hat{\mathbf{g}}_u$ is the uniform and $\hat{\mathbf{g}}_s$ the staggered component.\cite{Herak-11} \\
\indent To model the experimentally observed AFMR modes, we first transform the Hamiltonian [\eqref{eq:Hamiltparts}] into magnetic free energy $\mathcal{F}$ per Cu site by applying the molecular field approximation.\cite{Rohrer-69} Neutron diffraction suggests four magnetic sublattices so we write our $\mathcal{F}$ as (see \fref{fig12}), 
\begin{align}\label{eq:freeF}
\nonumber
	\mathcal{F} &= 2\sum\limits_{i=1}^2 \big(J'\: \mathbf{M}_{2i-1}\cdot \mathbf{M}_{2i} + J_{IC}'\: \mathbf{M}_i \cdot \mathbf{M}_{i+2}\: + \\\nonumber
	&+ \:\mathbf{D'} \cdot \mathbf{M}_{2i-1} \times \mathbf{M}_{2i}\:  +  \mathbf{M}_{2i-1}\cdot \hat{\bm{\delta}} J' \cdot \mathbf{M}_{2i}\big)\: -\\
	&- \sum\limits_{i=1}^4 \mathbf{M}_i \cdot (\hat{\mathbf{g}}_u - (-1)^i)\hat{\mathbf{g}}_s)/g_0 \cdot \mathbf{H} \: ,
\end{align}
where the factor 2 emerges from the boundary conditions. \eqref{eq:freeF} represents the simplest expression in which each spin has two intrachain neighbors and two interchain neighbors (see \fref{fig12}), which seems to be a good approximation for \csos, as mentioned above. In \eqref{eq:freeF} the staggered DM interaction and the staggered $g$ tensor are taken into account. The sublattice magnetizations are given by $\mathbf{M}_i=-N g \mu_B \left< \mathbf{S}_i \right>$, where $N$ is the number of Cu$^{2+}$ ions on the $i$-th sublattice, $g_0=2.0023$ is the free-electron $g$ factor and $\left< ... \right>$ indicates the thermal averaging. The relation between molecular-field constants and the interaction constants of the Hamiltonian \eref{eq:Hamiltonian} are defined by
%
\begin{subequations}
\begin{align}\label{eq:mfJ}
J' &= \dfrac{J}{N (g\mu_B)^2}\: ,\\
J_{IC}' &= \dfrac{J_{IC}}{N (g\mu_B)^2}\: ,\\
\mathbf{D}' &= \dfrac{1}{N (g\mu_B)^2}\: \mathbf{D}\: .
\end{align}
\end{subequations}
\indent The ground state of the system described by the expression \eref{eq:freeF} is obtained by numerical minimization of the free energy. Using the parameters suggested previously from the ESR analysis in the paramagnetic phase, $J=157$~K, $J_{IC}=0.1~J$, $|D_a^*|=0.044$, $|D_c|=0.0255$, $\delta_{a^*}=\delta_b=0$, $|\delta_c|=0.04$\cite{Herak-11}, and taking $\mu_{eff}=0.5\mu_B$ from the neutron scattering data we correctly predict the magnetic structure if $\delta_c$ is negative.  The effective field acting on the $i$-th sublattice magnetization is calculated from $\mathbf{B}_{eff,i}=-\partial \mathcal{F}/\partial \mathbf{M}_i$. The AFMR modes are then obtained from the equation of motion for $\mathbf{M}_i$ precessing around $\mathbf{B}_{eff,i}$.\cite{Pregelj-07} The four sublattice model predicts four AFMR modes, where only the lowest in energy is experimentally observed. We note that our previous EPR linewidth analysis\cite{Herak-11} could not determine the sign of the anisotropy parameter. However, these parameters fail to predict the precise value of the SF field and some details of the AFMR modes. \\
\indent We have managed to describe both the field as well as the angular dependence of the AFMR data by tuning only the symmetric anisotropic exchange parameters; $\delta_{a^*}=0.00046$, $\delta_b=0$, $\delta_c=-0.001$. The good agreement with the experimental results is demonstrated in Figs. \ref{fig9} and \ref{fig10}. We note that our calculations predict two modes (solid and dashed lines in \fref{fig9}), while only one mode was observed in measurements. The reason behind is that the second mode is very flat and hence extremely difficult to detect. \\
\indent Moreover, the same set of parameters describes also the magnetization measurement, as can be seen in \fref{fig2}(b). The values of magnetization obtained by calculation were rescaled to match the measured values. The spin-flop transition is observed for the $b$ direction at the field of $H_{SF}\approx 12.5$~kOe, in perfect agreement with the experiment. The sublattice magnetizations in the $(abc)$ coordinate system obtained for these parameters are $\bm{M}_{1,2}=(-0.0069, \pm 0.4998, -0.0136)$, $\bm{M}_{3,4}=(0.0069, \mp 0.4998, 0.0136)$, which corroborates with $b$ being the easy axis, as well as a finite magnetization of a single chain. Furthermore, our model also correctly predicts the magnetic order above the spin-flop transition. At $H=40$~kOe our calculations yield $|m_{a^*}|=0.46$ and $|m_c|=0.19$, in good agreement with the neutron experiment.
\\
\indent Finally, we also calculated angular dependence of the torque in the $a^*b$ plane. The results are shown with full red lines in \fref{fig11}. The amplitude of the measured torque depends on the mass of the sample, so calculated torque was rescaled to match the observed torque amplitude in the same way as magnetization. The agreement is excellent. Since the modeling with the experimental $\Delta\chi$ does not work, the complementary free-energy approach clearly demonstrates that the spins progressively rotate away from the $b$ axis in the $a^*c$ plane with rotation of the applied field. 
\section{\label{sec:disc}Discussion}
\indent Magnetic structure, AFMR, magnetization and angular dependence of the magnetic torque were correctly reproduced using the Hamiltonian proposed in the previous ESR study of the PM state.\cite{Herak-11} Present experiments allow for some further improvement. First, in the previous ESR study it was impossible to determine the sign of $\delta_c$. The free energy analysis presented here showed that in order to reproduce the correct ground state in agreement with the neutron diffraction and $\mu$SR measurements, it is necessary to have $\delta_c<0$. Furthermore, we were able to precisely assess the values of the symmetric anisotropic exchange, which was in the previous study least defined. The new values $\delta_{a^*}=0.00046$, $\delta_b=0$ and $\delta_c=-0.001$ are significantly smaller than previously proposed $\delta_{a^*}=\delta_b=0$, $\delta_c=-0.04$. The most likely reason for the discrepancy is that in the previous ESR study this parameter was derived based on the assumption that the observed linear temperature increase of the ESR linewidth results solely from the 1D spin-spin correlations below $T\leq 100$~K,\cite{Choi-11} as suggested by theory for 1D HAF.\cite{OA-PRL99,*OA-02,*OA-07, Maeda-05} In fact, part of the observed linear behavior might as well arise from the spin-phonon line broadening, which was observed at higher temperatures. \\
\indent The magnitude of the ordered Cu$^{2+}$ magnetic moment obtained from the present measurements is $\mu_{eff}\approx 0.5 \mu_B$. The moment is thus significantly reduced with respect to $1\mu_B$. The measured value is in good agreement with the predictions of the coupled quantum spin-chain approach, which gives $\mu_{eff}=0.46\mu_B$ for $J_{ic}=0.1\;J$.\cite{Rosner-97} The present results, similarly as previous ESR analysis, thus point to strong quantum spin fluctuations as anticipated in the quasi-1D systems. This conclusion opposes the conclusions drawn from Raman scattering, which indicate more classical 3D spin dynamics.\cite{Choi-11} \\
\indent Our susceptibility measurements along the three crystal directions allowed us to determine the anisotropic staggered field coefficient in \csos, which is a result of both the staggered $g$ tensor and DMI. The derived staggered field coefficient $c_c$=0.15(2) is comparable to those found in related Cu-based 1D compounds [PM·Cu(NO$_3$)$_2$·(H$_2$O)$_2$]$_n$ ($c_{c"}\approx 0.25$, $c_b \approx 0.16$)\cite{Feyerherm-00} and Cu benzoate ($c_c=0.20$).\cite{Dender-97} The staggered $g$ tensor and the observed DMI in \cso place this system to a subclass of quasi-1D spin systems where a staggered field can be induced by a magnetic field. The frequency span accessible in present experiments, however, seems not sufficient to directly probe the low-energy excitations characteristic of decoupled TLL--chains. In addition, strong staggered fields may drive \cso further away from the TTL state explaining why only spin-waves were observed as low-energy excitations. The crystal symmetry and the ordered state found in \cso are such that the competitive case should be realized when a finite magnetic field is applied below $T_N$. The staggered field coefficient found in \cso amounts to $c_c \approx 0.15(2)$, while the interchain interaction amounts to $J_{IC}/k_B\approx 20$~K ($1$~K $\sim 7.4$~kOe) which means that a very high external magnetic field $H\sim 1000$~kOe should be employed for \cso to induce the competition between the staggered field $h=cH$ and the staggered field originating from LRO structure. The observed spin-flop transition at $H_{SF}\approx 13$~kOe applied along the easy axis direction thus presents a classical spin-flop transition which originates from the competition of underlying anisotropies and is not driven by the staggered field, such as the spin reorientation transitions observed in BaCu$_2$Si$_2$O$_7$.\cite{Glazkov-05} We note that the latter is characterized by a similar staggered field coefficients but with much smaller interchain interaction, on the other hand.\cite{Casola-12} \\
\indent The observed decrease of susceptibility [inset of \fref{fig2}(a)] and susceptibility anisotropy\cite{Herak-11} below $22$~K which cannot be explained in 1D HAF model even when including staggered fields deserves a comment, albeit of a speculative nature. It is possible that this is connected to the crossover from 1D to 3D behavior of a 1D HAF in a staggered field. Recent NMR results on BaCu$_2$Si$_2$O$_7$ were satisfactorily explained using a Ginzburg-Landau free energy expansion in the vicinity of $T_N$ where this crossover was revealed.\cite{Casola-12} Further investigations of the magnetic response of \cso in the vicinity of $T_N$ by using other local probes, such as nuclear magnetic resonance, should prove very informative in this respect.
\section{\label{sec:concl}Conclusion}
\indent The magnetic ground state and the low-energy excitations in quasi-1D HAF \cso were studied experimentally by the neutron diffraction, static magnetic measurements, $\mu$SR and AFMR. All experimental results were coherently explained with the same Hamiltonian as derived previous from the analysis of the EPR linewidth within the theory for 1D HAF.\cite{Herak-11} The antiferromagnetically ordered ground state below $T_N=17$~K is characterized by the reduced Cu$^{2+}$ ($S=1/2$)magnetic moment 0.52(2)$\mu_B$, which is in line with the expected strong quantum fluctuations emerging from the underlying one-dimensionality of the system. Staggered magnetic fields arising from the staggered $g$ tensor and DMI govern the ground state and low-energy magnetic properties of the system, however, within experimentally accessible magnetic field they are too small to prevail over the interchain interaction and thus induce the TLL physics.  Nevertheless, future studies of this system in the vicinity of the phase transition could provide intriguing new insight about the influence of the staggered fields on a dimensional crossover, expected in quasi-1D systems with long-range order.
\begin{acknowledgments}
M. H. acknowledges financial support by the Postdoc program of the Croatian
Science Foundation (Grant No. O-191-2011), the Slovene
Human Resources Development and Scholarship fund under
grant No. 11013-57/2010-5 and the Croatian Ministry of Science, Education and Sports under Grant No. 035-0352843-2846. A. Z., M. P. and D. A.
acknowledge the financial support of the Slovenian Research
Agency (projects J1-2118 and BI-US/09-12-040). Neutron diffraction experiments were performed at SINQ, Paul Scherrer Institute, Villigen, Switzerland.
\end{acknowledgments}
\appendix
\section{\texorpdfstring{Comparison of calculated and observed intensities in neutron diffraction for $\Gamma_1$ and $\Gamma_3$ models of magnetic structure in \cso}{Comparison of calculated and observed intensities in neutron diffraction for Gamma1 and Gamma3 model of magnetic structure in CuSe2O5}}\label{appA}
\indent In \fref{fig3} the agreement of the two possible models, $\Gamma_1$ and $\Gamma_3$ (see Tab. \ref{tab:irreps}), with the observed intensities, is shown. In Table \ref{tab:neutrint} we list the magnetic intensities measured at 6~K and compare them to the calculated intensities for the $\Gamma_1$ and $\Gamma_3$ model. 
\begin{table}[tb]
	\centering
		\caption{Observed and calculated magnetic intensities $I_{obs}$ and $I_{calc}$, respectively, of \cso single crystal at 6~K corresponding to the models $\Gamma_1$ and $\Gamma_3$ discussed in the text.}
		\begin{tabular*}{\linewidth}{@{\extracolsep{\fill}}  r  r  r  r  r  r }
		\hline \hline
	 $h$ & $k$ & $l$ & $I_{obs}$  &  $I_{calc}$ ($\Gamma_1$) & $I_{calc}$ ($\Gamma_3$) \\\hline
   1 &  0 &  0 &   2.00 & 3.17 & 1.66   \\
  -1 &  0 &  1 & 172.42 & 162.88 & 156.87 \\
   1 &  0 &  1 & 140.45 & 115.9 &  146.33 \\             
   0 & -1 &  0 &   13.48 & 0 &   9.52 \\
  -3 &  0 &  1 &  108.31 & 22.27 & 133.15 \\          
  -1 &  0 &  2 &    9.01 &  2.56 &  7.77 \\
   2 & -1 &  0 &   11.88 &  1.05 &  5.44 \\
   2 & -1 &  1 &   63.93 &  88.95 & 68.97 \\
   3 &  0 &  1 &   73.92 &  38.41 & 109.25 \\
  -1 &  0 &  3 &  145.91 &  123.63 & 96.95 \\
  -3 &  0 &  3 &  100.76 &  98.27 & 94.64 \\
  -5 &  0 &  1 &   53.88 &  2.5 & 91.67 \\
   1 &  2 &  1 &   17.85 &  96.27 & 14.19 \\
  -5 &  0 &  3 &   64.05 &  45.59 & 75.13 \\
  -1 &  0 &  5 &   64.12 &  49.88 & 39.46 \\
   2 &  1 & -1 &   52.95 &  116.69 &  54.53 \\
   0 & -3 &  0 &    5.00 &  0 &  3.10 \\
   2 &  1 &  0 &    2.97 &  1.05 &  5.44 \\
   1 &  2 &  0 &    4.24 &  0.08 &  5.70 \\
   0 & -1 &  2 &   10.92 &   1.36 & 7.23 \\ \hline \hline
		\end{tabular*}
		\label{tab:neutrint}
\end{table}
\section{\texorpdfstring{Muon polarization in the ${\rm CuSe_2O_5}$~single crystal}{Muon polarization in the CuSe2O5 single crystal}}\label{appB}
\label{ap}
We set the orthogonal coordinate system so that its $x$, $y$ and $z$ axes corresponds to the right, up and backward direction with respect to the muon-beam direction, respectively. In general, the initial muon polarization is tilted by an angle $\alpha$ from the $z$ axis in the $yz$ plane, ${\bf P}_0=(0,\,{\rm sin}\alpha,\,{\rm cos}\alpha)$, and the local field is characterized by the polar angle $\theta$ and the azimuthal angle $\varphi$, ${\bf B}_\mu=B_\mu({\rm sin}\theta\,{\rm cos}\phi,\,{\rm sin}\theta\,{\rm sin}\phi,\,{\rm cos}\theta)$. The parallel and the perpendicular component of the muon polarization with respect to the field are then derived from the equation of motion ${\rm d}{\bf P}(t)/{\rm d} t = \gamma_\mu {\bf P}(t) \times {\bf B}_\mu$,
\begin{eqnarray}
{\bf P}^\| &=& \left({\bf P}_0\cdot{\bf B}_\mu\right)\frac{{\bf B}_\mu}{B_\mu},\\
{\bf P}^\bot(t) &=& {\bf P}^\bot_{\rm 1}{\rm cos}(\gamma_\mu B_\mu t) + {\bf P}^\bot_{\rm 2}{\rm sin}(\gamma_\mu B_\mu t),
\label{eqa1}
\end{eqnarray}
respectively, where the two orthogonal perpendicular vectors are 
\begin{eqnarray}
{\bf P}^\bot_{\rm 1} &=& {\bf P}_0-{\bf P}^\|,\\
{\bf P}^\bot_{\rm 2} &=& {\bf P}^\bot_{\rm 1}\times \frac{{\bf B}_\mu}{B_\mu}.
\label{eqa2}
\end{eqnarray}
If the longitudinal muon relaxation rate $\lambda_L$ and the transverse relaxation rate $\lambda_T$ are taken into account, the total muon polarization at a given time will be given by
\begin{equation}
{\bf P}(t) = {\bf P}^\| {\rm e}^{-\lambda_L t} +  {\bf P}^\bot (t) {\rm e}^{-\lambda_T t}
\label{eqa3}
\end{equation}
The muon polarization measured by the backward-forward and the up-down sets of detectors is then changing with time as
\begin{eqnarray}
P^z (t,B_\mu,\theta,\varphi,\lambda_L,\lambda_T,\alpha) &=& {\bf P}(t)\cdot (0,0,1),\\
P^y (t,B_\mu,\theta,\varphi,\lambda_L,\lambda_T,\alpha) &=& {\bf P}(t)\cdot (0,1,0),
\label{eqa4}
\end{eqnarray}
respectively.

In the \cso~single crystal each muon stopping site~$i$ gives four different orientations of the given magnetic field, because of the symmetry of the magnetic little group (only the inversion symmetry leaves the field unchanged). Therefore, the polarizations are given by
\begin{eqnarray}
P^{y,z}_{\rm sc} (t,B_{\mu i},\theta_i,\varphi_i,\lambda_{L i},\lambda_{T i},\alpha)= \nonumber \\
\frac{1}{4}\sum_{j=1}^4{P^{y,z} (t,B_{\mu i},\theta_i^j,\varphi_i^j,\lambda_{L i},\lambda_{T i},\alpha)}.
\label{eqa5}
\end{eqnarray}
The four polar and azimuthal angles for each site are symmetry related. For the first orientation of the crystal with the $a^*$ crystallographic axis along $z$ direction and the $b$ axis along $y$ direction, these relations are
\begin{eqnarray}
\theta_i^1 = \theta_i&,&\;\varphi_i^1 = \varphi_i \nonumber\\
\theta_i^2 = \theta_i&,&\;\varphi_i^2 = - \varphi_i \nonumber\\
\theta_i^3 = \pi-\theta_i&,&\;\varphi_i^3 = \pi+\varphi_i \nonumber\\
\theta_i^4 = \pi-\theta_i&,&\;\varphi_i^4 = -\pi-\varphi_i.
\label{eqa6}
\end{eqnarray}
In the second orientation the crystal is rotated by $\pi/2$ around the $z$ axis, which leaves $\theta_i^j$ unchanged and changed $\varphi_i^j\rightarrow\varphi_i^j+\pi/2$.
%

%
%
%
\end{document}